*Full Length Research Paper*

# Application of Bousinesq's and Westergaard's formulae in analysing foundation stress distribution for a failed telecommunication mast

OJEDOKUN Olalekan Yinka[1]* and OLUTOGE Festus Adeyemi[2]

[1]Department of Civil Engineering, Faculty of Engineering, the Polytechnic, Ibadan, Oyo State, Nigeria.
[2]Department of Civil Engineering, Faculty of Engineering, University of Ibadan, Oyo State, Nigeria.

The concurrent foundation failure of telecommunication masts in Nigeria and all over the world which endanger the lives and properties of residents situated within the fall distance of the telecommunication mast is a thing of great concern. In this study, a GSM mast that underwent foundation failure at Ibadan, Oyo State, Nigeria was critically examined with a view to providing engineering solution. The soil investigation at the global system for mobile communications (GSM) telecommunication tower comprised of laboratory tests: sieve analysis, Atterberg limits and moisture content tests were carried out on the soil samples obtained while Dutch cone penetrometer test was performed on the site to a depth of refusal to determine the allowable bearing pressure at various depths of the soil. The application of Boussineq's and Westergard's formulae for point loads using Java programme to simulate and compute the stress distribution at various predetermined depths showed the stress distribution pattern beneath the failed foundation of the structure. The stress distribution pattern revealed that the soil strength was lower than the imposed loadings from the structure thereby resulting in differential settlements and cracks at the foundation. A variety of engineering solutions were recommended to improve the soil strength and thus prevent such occurrences in future.

## INTRODUCTION

A telecommunication mast installation comprises of a mast supporting telecommunications antenna and a foundation structure supporting the mast, the foundation structure being in the form of an enclosed chamber situated at least partially underground and defining an internal space which is accessible to personnel and which accommodates electronic equipment associated with operation of the antenna. The telecommunication mast has a foot at its lower end which is supported on a base of the chamber, the base acting as a structural foundation for the mast. The foot of the mast is seated on the base, the seat restraining lateral movements of the foot of the mast at the base without transfer of bending moments between the mast and the foundation structure (Creighton, 2002).

Numerous investigations have been carried out on the design and erection techniques of telecommunication towers. Comparatively little attention has been directed toward the behaviour and deterioration of tower foundations. It should be pointed out that the design of tower foundations is more involved than that of other steel structures. In the latter case, the foundations are usually subjected to static compressive force with uniform stress distribution on soil (Abdalla, 2002).

Communication masts are used for all types of wireless communication and come in many different shapes and sizes. From a structural point of view, a lattice construction has been shown to provide a strong durable structure upon which to locate antennae of different types and size. However, many people consider lattice constructions to be unattractive and in the recent past this has resulted in slim tubular constructions being utilized in order to reduce the visual impact of the mast (Heslop, 2002).

---

*Corresponding author. E-mail: daniely2kus@yahoo.com.

In an engineering sense, failure may occur long before the ultimate load or the load at which the bearing resistance of the soil is fully mobilized since the settlement will have exceeded tolerable limits. Terzaghi (1967) suggested that, for practical purposes, the ultimate load can be defined as that which causes a settlement of one-tenth of the pile diameter or width which is widely accepted by engineers. In most cases where the piles are acting as structural foundations, the allowable load is governed solely from considerations of tolerable settlement at the working load. An ideal method of calculating allowable loads on piles would be one which would enable the engineer to predict the load-settlement relationship up to the point of failure, for any given type and size of pile in any soil or rock conditions (Tomlinson, 1994).

Estimation of vertical stresses at any point in a soil-mass due to external vertical loadings is of great significance in the prediction of settlements of buildings, bridges, embankments and many other structures. Equations have been developed to compute stresses at any point in a soil mass on the basis of the theory of elasticity. According to elastic theory, constant ratios exist between stresses and strains. For the theory to be applicable, the real requirement is not that the material necessarily be elastic, but there must be constant ratios between stresses and the corresponding strains. Therefore, in non-elastic soil masses, the elastic theory may be assumed to hold so long as the stresses induced in the soil mass are relatively small. Since the stresses in the subsoil of a structure having adequate factor of safety against shear failure are relatively small in comparison with the ultimate strength of the material, the soil may be assumed to behave elastically under such stresses. When a load is applied to the soil surface, it increases the vertical stresses within the soil mass. The increased stresses are greatly directly under the loaded area, but extend indefinitely in all directions. Many formulae based on the theory of elasticity have been used to compute stresses in soils. They are all similar and differ only in the assumptions made to represent the elastic conditions of the soil mass. The formulae that are most widely used are the Boussinesq and Westergaard formulas (Murthy, 1992).

**METHODOLOGY**

**Collection of samples**

The soil investigation at the GSM telecommunication tower at Ibadan, Oyo State, Nigeria comprised of and was carried out in three parts; field work: (tests two boreholes), laboratory analysis of borehole samples obtained and analysis of the test results. The scope of work executed involved the performance of two boreholes to depth of refusal. The samples obtained from the borehole test were also subjected to laboratory analysis. The laboratory tests carried out on the samples are: grain size analysis, moisture content, atterberg limits, dutch cone penetrometer tests.

**Pile details**

Mass of tower ad ladder, W  = 12796 Kg
Height of tower, Ht             = 55 m
Maximum force/leg, Rv         = 524.43 kN
Maximum uplift/leg, Wv        = -461.67 kN
Leg spacing heal-heal          = 6502 mm
R.C. Stud section               = 700 × 700 mm
Load factor                       = 1.3

**Safe working loads**

Pile depth, Lp = -6.00 m

**Pile requirement**

Piles required/leg = Max forces per leg/(Qs + Qb)

300 mm        = 524.43/108 = 4.8558333 ~ 5No
400 mm        = 524.43/182 = 2.8814835 ~ 3No

Provision per leg
400 mm        = 3No

**Load per pile**

Wp = 174.81 kN
Wp (factored) = 174.81 x 1.3 = 227.253 KN

**Pile design**

Designed as a short braced column min. steel required = 0.4%ACol

$A_{col}$ = (3.142 × $400^2$) / 4  = 125680 $mm^2$

$A_{st}$ = (0.4 × 125680) / 100 = 502.72 $mm^2$

Provide = 6 No. 16 mm diameter bar

Asc(pro) = 1206 $mm^2$

Load capacity N = 0.35 Fcu x Acol + 0.67Ast x Fy
           N = 0.35 x 25 x 125680 + 0.67 x 1206 x 380
           N = 1406.70kN

**RESULTS AND DISCUSSION**

**Physical properties of soil samples**

The analysis of total load to be carried by the piles is presented in Table 1. The results of the tests carried out on the soil specimens are given in Table 2. Sample 4 had the highest value of plastic and liquid limits while samples 1, 3, 5, 7 and 8 had no plastic and liquid limits. Generally, liquid limits vary widely, but values of 40 to 60% and above are typical of clay soils and values of 25 to 50% can be expected for silty soil so clays soils with liquid limits between 30 and 50% exhibit medium plasticity while liquid limits less than 30% infer low plasticity and liquid limits greater than 50% indicate high plasticity. High liquid limit is an indication of soils with high clay content and low bearing capacity.

Sample 8 had the highest moisture content which indicated poor drainage property while sample 1 had the lowest moisture content indicating that the drainage property is good at that depth.

**Allowable bearing pressure at predetermined depths**

The allowable bearing capacity at predetermined depths were determined for borehole 1 and tabulated in Table 3.

The highest bearing capacity of the soil which was at 7.50 m depth was 220 kN/m$^2$. The derived allowable bearing capacity of the soil in relationship to the actual load of the telecommunication mast would be used in the simulation using Java programme to determine the cause of failure.

**Application of Bousinesq's formula for stress distribution**

$$\sigma_B = \frac{3N}{2 \Pi Z^2} \frac{1}{[1 + (r/Z)^2]^{5/2}} = \frac{N I_B}{Z^2}$$

$$I_B = \frac{3}{2 \Pi} \frac{1}{[1 + (r/Z)^2]^{5/2}}$$

Where $\sigma_B$ = Bousinesq stress coefficient
$I_B$ = Bousinesq vertical stress
N = Point load from the end bearing pile
Z = Vertical distance from the end point of pile
r = The radial distance from Z (Murthy, 1992).

**Application of Westergaard's formula for stress distribution**

$$\sigma_W = \frac{N}{\Pi Z^2} \frac{1}{[1 + 2(r/Z)^2]^{3/2}} = \frac{N I_W}{Z^2}$$

$$I_W = \frac{(1/\Pi)}{[1 + 2(r/Z)^2]^{3/2}}$$

Where $\sigma_W$ = Westergaard stress coefficient
$I_W$ = Westergaard vertical stress
N = Point load from the end bearing pile
Z = Vertical distance from the end point of pile
r = Radial distance from Z (Murthy, 1992).

**Stress distribution at predetermined depths**

Based on the Bousinesq's and Westergaard's formulae stated, the Java coding for determining the various stresses at predetermined depths was used to analyse the experimental data. The result of the analysis is shown in Table 4.

Figure 1: Graph of values of $I_B$ and $I_W$ showed the values obtained from the Table 4: Stress distribution at predetermined depths when $I_B$ and $I_W$ were plotted against r/z. The graph (Figure 1) showed that both Bousinesq vertical stress and Westergaard vertical stress decreased as the depths moves further away from the point load. In other words, as the values of radial and vertical distances increased, both values of Bousinesq vertical stress and Westergaard vertical stress decreased.

Figure 2: Graph of values of $\sigma_B$ and $\sigma_W$ showed the values obtained from the Table 3: Stress distribution at predetermined depths when $\sigma_B$ and $\sigma_W$ were plotted against r/z. The graph (Figure 2) showed that both Bousinesq vertical stress coefficient and Westergaard vertical stress coefficient decreased as the depths moved further away from the point load. In other words, as the values of radial and vertical distances increased, both values of Bousinesq Vertical Stress and Westergaard Vertical Stress decreased.

Figure 3: Graph of values of $\sigma_B/\sigma_W$ showed the values obtained from the Table 4: Stress distribution at predetermined depths when $\sigma_B/\sigma_W$ is plotted against r/z. The graph (Figure 3) showed that $\sigma_B/\sigma_W$ decreased as the depths moved further away from the point load. In other words, as the values of radial and vertical distances increased, $\sigma_B/\sigma_W$ decreased.

**Conclusion**

From the analysis of the results obtained from *in-situ* and laboratory tests carried out on the soil samples at the telecommunication mast site, the following conclusions can be drawn.

(1) The telecommunication mast was erected at a distance of two metres to the nearby building thereby violating the Nigerian Communication Commission guidelines of observing a clear fall distance to any nearby building.

The stress distribution at predetermined depths revealed:

**Table 1.** Load classification on piles.

| Pile diameter (mm) | End bearing, Qb (kN) | Skin friction, Qs (kN) | Total load {Qb (kN) +Qs(kN)} |
|---|---|---|---|
| 300 | 84 | 24 | 108 |
| 400 | 150 | 32 | 182 |

**Table 2.** Physical properties of soil samples.

| BH number | Sample number | Depth (m) | Soil type | Natural moisture content (%) | Atterberg limits (%) | | | Grading analysis percentage passing | | | | | | | | | |
|---|---|---|---|---|---|---|---|---|---|---|---|---|---|---|---|---|---|
| | | | | | LL | PL | PI | 9.5 | 6.3 | 4.75 | 2.36 | 1.18 | 600 | 425 | 300 | 212 | 150 | 75 |
| 1 | 1 | 0.15 | SM | 13 | | | | | | | 100 | 98 | 89 | 80 | 75 | 61 | 48 | 34 |
| | 2 | 2.25 | SC | 17 | 44 | 18 | 26 | | | | 100 | 97 | 88 | 80 | 71 | 65 | 55 | 48 |
| | 3 | 3.75 | SC | 16 | | | | | | | 100 | 97 | 88 | 80 | 71 | 61 | 54 | 47 |
| | 4 | 5.25 | SC | 19 | 46 | 19 | 27 | | | | 100 | 98 | 92 | 85 | 77 | 67 | 58 | 49 |
| | 5 | 7.50 | GP/SM | 20 | | | | 85 | 81 | 80 | 72 | 67 | 56 | 48 | 41 | 34 | 29 | 23 |
| 2 | 6 | 0.75 | SM/SC | 18 | 31 | 13 | 18 | 100 | 99 | 96 | 90 | 77 | 68 | 59 | 48 | 41 | 34 |
| | 7 | 3.00 | SM/SC | 18 | | | | | 100 | 98 | 94 | 81 | 69 | 58 | 47 | 39 | 32 |
| | 8 | 6.00 | SC | 23 | | | | 100 | 99 | 98 | 94 | 83 | 74 | 66 | 57 | 50 | 45 |
| | 9 | 8.25 | SC | 16 | 45 | 17 | 28 | | 100 | 99 | 93 | 81 | 73 | 66 | 58 | 53 | 48 |

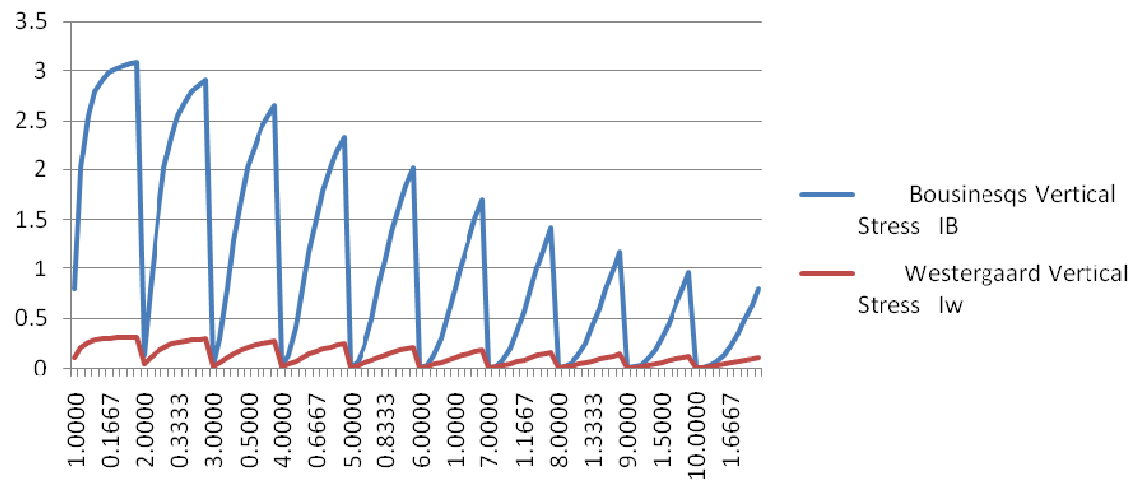

**Figure 1.** Graph of $I_B$ and $I_W$.

**Table 3.** Allowable bearing pressure at predetermined depths.

| Depth (m) | Allowable bearing pressure (kN/m$^2$) |
|---|---|
| 0.00 - 0.15 | 50 |
| 0.15 - 2.25 | 108 |
| 2.25 - 3.75 | 140 |
| 3.75 - 5.25 | 190 |
| 5.25 - 7.50 | 220 |

**Table 4.** Stress distribution at predetermined depths based on Java coding.

| Radial distance, r(m) | Vertical distance z (m) | r/z | Bousinesq's vertical stress ($I_B$) | Bousnesq's stress coefficient, $\sigma_B$(N/m$^2$) | Westergaard vertical stress ($I_w$) | Westergaard stress coefficient, $\sigma_w$ (N/m$^2$) | $\sigma_B/\sigma_w$ |
|---|---|---|---|---|---|---|---|
| 1 | 1 | 1.0000 | 0.7854 | 1104745.006 | 0.1061 | 149245.4289 | 7.4022 |
| 1 | 2 | 0.5000 | 2.0106 | 2828147.216 | 0.2122 | 298490.8577 | 9.4748 |
| 1 | 3 | 0.3333 | 2.5447 | 3579373.82 | 0.2604 | 366329.689 | 9.7709 |
| 1 | 4 | 0.2500 | 2.7829 | 3914390.61 | 0.2829 | 397987.8103 | 9.8355 |
| 1 | 5 | 0.2000 | 2.9046 | 4085595.437 | 0.2947 | 414570.6357 | 9.855 |
| 1 | 6 | 0.1667 | 2.9741 | 4183344.129 | 0.3016 | 424171.2189 | 9.8624 |
| 1 | 7 | 0.1429 | 3.0172 | 4243988.416 | 0.3058 | 430178.0009 | 9.8657 |
| 1 | 8 | 0.1250 | 3.0457 | 4284057.321 | 0.3087 | 434168.5203 | 9.8673 |
| 1 | 9 | 0.1111 | 3.0654 | 4311857.219 | 0.3106 | 436947.4604 | 9.8681 |
| 1 | 10 | 0.1000 | 3.0797 | 4331908.661 | 0.3121 | 438957.1437 | 9.8686 |
| 2 | 1 | 2.0000 | 0.1257 | 176759.201 | 0.0354 | 49748.4763 | 3.5531 |
| 2 | 2 | 1.0000 | 0.7854 | 1104745.006 | 0.1061 | 149245.4289 | 7.4022 |
| 2 | 3 | 0.6667 | 1.5057 | 2117972.675 | 0.1685 | 237036.8576 | 8.9352 |
| 2 | 4 | 0.5000 | 2.0106 | 2828147.216 | 0.2122 | 298490.8577 | 9.4748 |
| 2 | 5 | 0.4000 | 2.3347 | 3284022.016 | 0.2411 | 339194.1565 | 9.6818 |
| 2 | 6 | 0.3333 | 2.5447 | 3579373.82 | 0.2604 | 366329.689 | 9.7709 |
| 2 | 7 | 0.2857 | 2.6853 | 3777134.582 | 0.2736 | 384896.106 | 9.8134 |
| 2 | 8 | 0.2500 | 2.7829 | 3914390.61 | 0.2829 | 397987.8103 | 9.8355 |
| 2 | 9 | 0.2222 | 2.8529 | 4012861.999 | 0.2897 | 407490.3283 | 9.8477 |
| 2 | 10 | 0.2000 | 2.9046 | 4085595.437 | 0.2947 | 414570.6357 | 9.855 |
| 3 | 1 | 3.0000 | 0.0314 | 44189.8002 | 0.0168 | 23565.0677 | 1.8752 |
| 3 | 2 | 1.5000 | 0.2974 | 418364.9728 | 0.0579 | 81406.5976 | 5.1392 |
| 3 | 3 | 1.0000 | 0.7854 | 1104745.006 | 0.1061 | 149245.4289 | 7.4022 |
| 3 | 4 | 0.7500 | 1.2868 | 1810014.218 | 0.1498 | 210699.429 | 8.5905 |

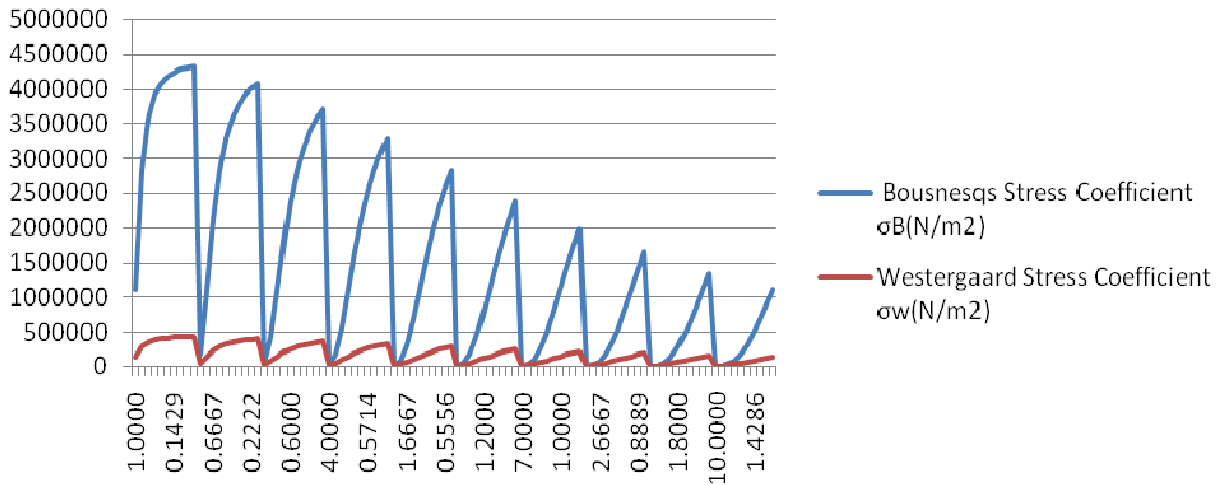

**Figure 2.** Graph of σ_B and σ_W.

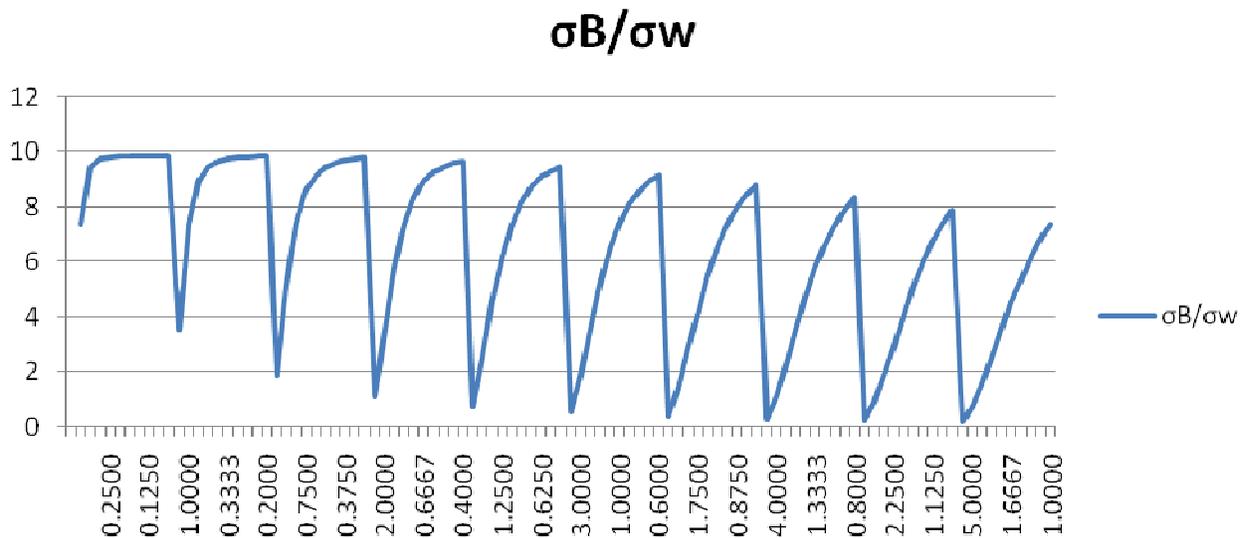

**Figure 3.** Graph of σ_B/σ_W.

1) Bousinesq vertical stress, Westergaard vertical stress, Bousinesq vertical stress coefficient and Westergaard vertical stress coefficient decreased as the depths moved further away from the point load. In other words, as the values of radial and vertical distances increased, the values of Bousinesq vertical stress, Westergaard vertical stress, Bousinesq vertical stress coefficient and Westergaard vertical stress coefficient decreased.
2) The total weight of the structure transmitted by the end bearing of the pile foundation unto the soil which is 389.47 kN/m$^2$ in relationship to the soil bearing capacity at the depth of the pile foundation which is 220 kN/m$^2$ has caused a differential settlement to occur at the foundation. This differential settlement has resulted to cracks occurring at the pile caps and also a clearance at the base plate which would eventually result to a total collapse if not attended to.

Based on the conclusions arrived at, the following recommendations could be adopted;

1) The telecommunication mast should be dismantled because it violated the Nigeria Communication Commission guidelines for safe erection of telecommunication mast of observing a clear fall distance.
2) The soil bearing pressure could be improved by adopting any of the following soil improvement methods: Application of vertical or wick drains, vacuum consolidation, cement deep mixing, vibroflotation techniques, application of geotextiles.